\documentclass[10pt,a4paper]{IEEEtran}
\usepackage{mathptmx}      
\usepackage{graphicx}
\usepackage{amssymb,amsmath}
\usepackage{breqn}
\usepackage{epstopdf}
\usepackage{cite}
\newtheorem{theorem}{Theorem}[section]
\newtheorem{lemma}[theorem]{Lemma}
\newtheorem{proposition}[theorem]{Proposition}

\begin{document}

\title{Analytical Coverage Probability of a Typical User In Heterogeneous Cellular Networks }
\author{Sinh Cong Lam        \and
Kumbesan Sandrasegaran  \\
Centre for Real-Time Information Networks, Faculty of Engineering and Information Technology, \\
University of Technology, Sydney, Australia
}

\maketitle

\begin{abstract}
In a Poisson Point Process (PPP) network model, in which the locations of Base Stations (BSs) are randomly distributed according to a Spatial Poisson Process,  has been recently used as a tractable stochastic model to analyse the performance of downlink Heterogeneous Cellular Networks (HCNs). The HCN is modelled as a multi-tier cellular network where each tier is characterised by the transmission power level, propagation path loss exponent and density of BSs. The current works on HCN enabling Intercell Interference Coordination (ICIC) technique usually deal with Strict Frequency Reuse (FR) or Soft FR with a reuse factor of $\Delta=1$ in a Rayleigh fading channel. It has been assumed that all Base Stations (BSs) transmit continuously  which leads to a reduction on the impact of number of users and RBs on network performance. In this paper, the performance of Soft FR with a reuse factor of $\Delta>1$ in Rayleigh-Lognormal fading channel is evaluated. The impact of the number of users and Resource Blocks (RBs) on Intercell Interference (ICI) are  presented for  Round Robin scheduling and indicator functions. The results show that there are opposite trends between coverage probability of Cell-Center User (CCU) and Cell-Edge User (CEU).

\end{abstract}

\section{Introduction}
In Orthogonal Frequency-Division Multiple Access (OFDMA) multi-cell networks, Intercell Interfere (ICI), which causes by the use of the same Resource Block (RB) at the same time at adjacent cells, has significant impact on network performance. Intercell Interference Coordination (ICIC)\cite{Softdefinition} such as Frequency Reuse (FR) has been introduced as a sufficient technique that can  mitigate the ICI then improve the performance of users having low Signal-to-Interference-plus-Noise Ratio (SINR). 

The Poisson Point Process (PPP) network model, which was presented in \cite{Andrews} for single-tier networks and then developed in \cite{ThomasDavidNovlan2011} for multi-tier networks, is considered as the accurate stochastic model to analyse the network performance. In single-tier PPP model, the number and locations of BSs are randomly generated according to a Poisson Distribution and Spatial Poisson Process respectively. Each BS covers a particular Voronoi cell. The PPP network model with K tiers is considered as the combination of K single-tier model in which each tier is characterised by the transmission power level, propagation path loss exponent and density of BSs.

Coverage probability approach was proposed to evaluate the performance of PPP network model \cite{Andrews,Dhillon2012,flexiblecell,optimalFrSpectrumwithoutnoise}. In \cite{Andrews} and \cite{Dhillon2012}, it was assumed that there was one user in each cell, hence the effects of scheduling algorithms on the network performance were not studied. In \cite{flexiblecell,optimalFrSpectrumwithoutnoise}, the Round Robin scheduling was considered, however the authors worked with a interference-limited network model (no noise). Furthermore, in these work, it has been assumed that the signal in downlink network only experiences fast fading, i.e. Rayleigh fading, and propagation pathloss, while in a realistic mobile radio network, the signal can be affected by slow fading, i.e. shadow fading, which  is caused by the variation in the terrain configuration between the transmitter and the receiver.

Some works that evaluated the effects of Rayleigh and shadowing
were considered in \cite{SinhCongLam2015,Keeler}. In \cite{Keeler}, shadowing was not incorporated in channel gain and can be constant when the origin PPP model is rescaled. In \cite{SinhCongLam2015}, the Rayleigh fading and Lognormal fading were considered together in the form of a Rayleigh-Lognormal channel and the closed-form expression of coverage probability was found by using Gauss-Legendre approximation.

In all papers discussed above, either Strict FR or Soft FR with reuse factor $\Delta=1$ was investigated that led to the fact that the all interfering BSs of a typical user in a given tier have the same transmission power and the same distribution densities. In addition, it has been assumed that all BSs transmit continuously which means that the adjacent cells always create ICI to the typical user. These assumptions significantly reduce the impact of the number of users and RBs on the network performance.

The main contribution of this work is to evaluate the coverage probability of a typical user in the random Poison network enabling Soft FR with factor $\Delta>1$. Since in Soft FR, the BSs at different cells in the same tier can transmit different power levels on the same RB, the transmission power ratio is defined to classify  these BSs. Another theoretical contribution is that the impact of the parameters including Round Robin scheduling, the number of users and RBs on the network performance are  considered. For this purpose, the indicator functions which present the probability where a BS causes ICI to a typical user are defined. A part of this work which analyse the performance of the Cell-Center User (CCU) in the single-tier network has been presented
in \cite{CongICC2016}. In this paper, the multi-tier network is investigated and the coverage probability of a CCU, Cell-Edge User (CEU) and random user that can be served as a CCU or CEU, are evaluated. The results of this paper differ from those in \cite{CongICC2016} because for multi-tier networks, the problem becomes more complex and more network parameters should be configured.

\section{System model}
In this paper, a PPP cellular network with K tiers in which the locations of BSs in tier $i$ are distributed as a homogeneous spatial Poisson Point Process (PPP) with density $\lambda_i$, $\left(0<i\leq K \right)$, is considered. The radio signals within the  tier $i$ experience path loss with exponent $\alpha_i$. Without loss of generality, a typical user is assumed to be located at the origin and served by the nearest BS in the tier $i$. 
The Probability Density Function (PDF) of the distance $r_i$ from a typical user to its nearest BS in tier $i$ is given by \cite{flexiblecell}
\begin{equation}
f_{R_i}(r_i)=\frac{2\pi\lambda_i}{P_{Ai}}r_i\exp\left(-\pi\sum\limits_{j=1}^{K}\lambda_jr^2 \right) 
\label{PDFdistance}
\end{equation}
in which $P_{Ai}$ is the probability that a typical user has a connection with the tier $i$.

\subsection{\large Fading channel model}
In realistic mobile radio scenarios in urban areas, the multipath effect at the mobile receiver due to scattering from local scatters such as buildings in the neighborhood of the receiver causes fast fading while the variation in the terrain configuration between the base-station and the mobile receiver causes  slow shadowing. The PDF of power gain $g$ of a signal experiencing Rayleigh and Lognormal fading is found from the PDF of the product two cascade channels \cite{DinhThiThai}
\begin{align*}
&f_{R-Ln}(g)\\
&=\frac{1}{\sigma_z\sqrt{2\pi}}\int\limits_{0}^{\infty}\frac{1}{x^2}\exp(-\frac{g}{x})\exp\left(-\frac{\left(10\log_{10}x-\mu_z\right)^2 }{2\sigma_z^2} \right)dx
\end{align*}
where $\mu_z$ and $\sigma_z$ are mean and variance of Rayleigh-Lognormal random variable (RV).
Then, its Cumulative Density Function (CDF) can be approximated by Gauss-Hermite quadrature as given below: \cite{SinhCongLam2015}
\begin{equation}
F_{R-Ln}(g)=\sum\limits_{n=1}^{N_H}\frac{\omega_n}{\sqrt{\pi}}\left[1-\exp(-\frac{g}{\gamma(a_n)}) \right] 
\end{equation}
in which $\gamma(a_n)=10^{(\sqrt{2}\sigma_za_n+\mu_z)/10}$; $w_n$  and $a_n$ are, respectively, the weights and the abscissas of the Gauss-Hermite polynomial.

The average of the power gain of Rayleigh-Lognormal channel is $\overline{g}_{R-Ln}=10^{(\mu_z+\frac{1}{2}\sigma_z^2)/10}$. In this paper, it is assumed that the power gain of the channel is normalised, i.e. $\overline{g}_{R-Ln}=1$.

\subsection{\large Frequency Reuse Algorithm}
\label{sec:Softpara}
We assume that all cells in a given tier use the same FR pattern including resource allocation and FR factor. For example, each cell in tier $i$ is allowed to use  $N_i$ resource blocks (RBs) which are divided into two groups  called cell-center and cell-edge RB group containing $N_{i}^{(c)}$ and $N_{i}^{(e)}$ RBs (such that $N_{i}^{(c)}+N_{i}^{(e)}=N_i$) respectively. Furthermore, $M_i$ users in each cell are classified  into $M_{i}^{(c)}$ Cell-Center Users (CCUs) and $M_{i}^{(e)}$ Cell-Edge Users (CEUs) (such that $M_{i}^{(c)}+M_{i}^{(e)}=M_i$) by SINR threshold $T_i$ . 
\begin{figure}[tbph]
\centering
\includegraphics[width=1\linewidth]{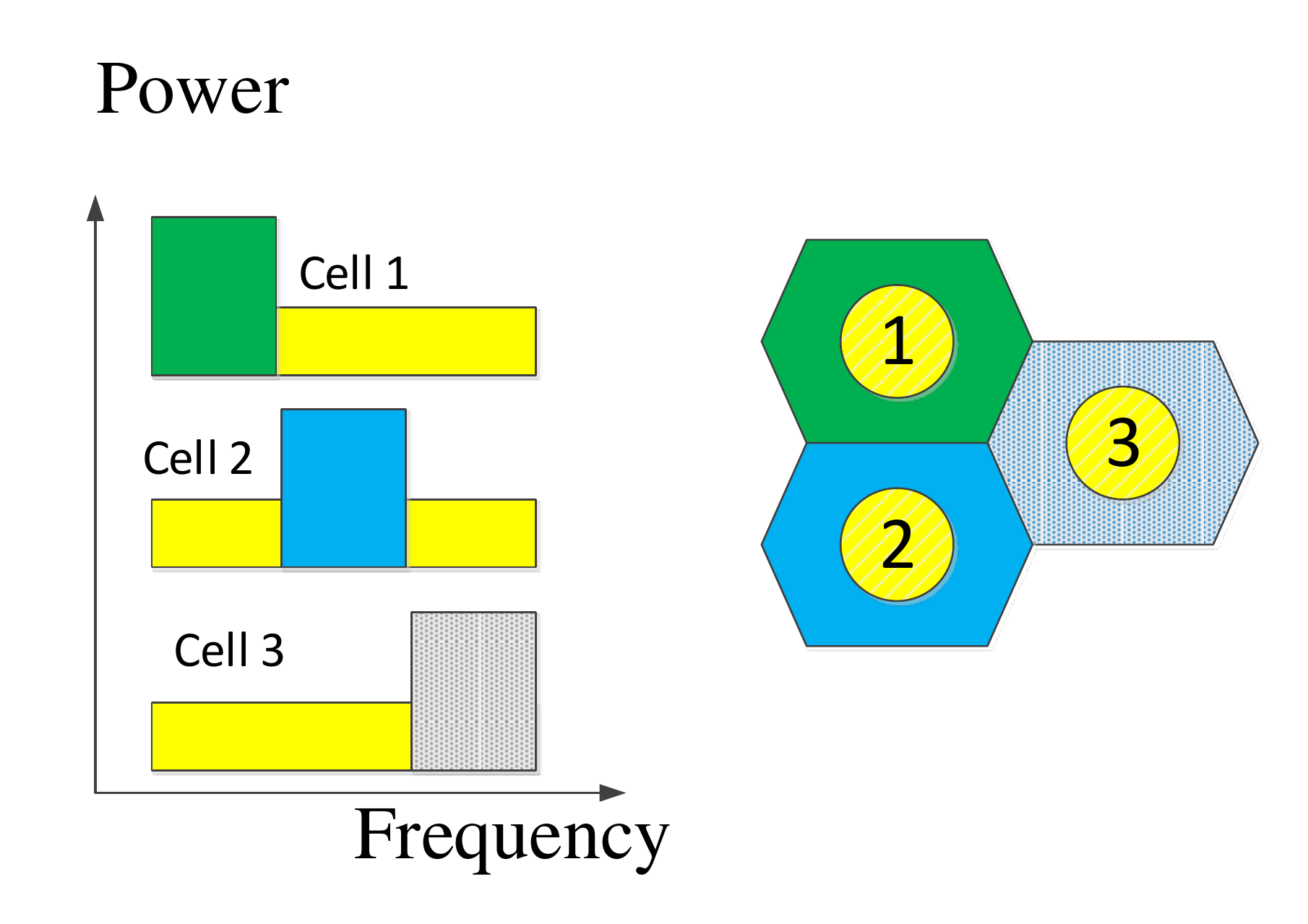}
\caption{Soft FR with $\Delta=3$}
\label{fig:SoftFR}
\end{figure}

We define $\phi_i$  to be  \textbf{transmission power ratio} on a cell-edge RB and cell-center RB, i.e. the transmission power on a cell-center and a cell-edge RB in tier $i$ are $P_i$ and $\phi_iP_i$. When the FR factor $\Delta_i$ is used, the densities of cells in the tier $i$, that uses the RB $b$  as a cell-center and cell-edge RB, are $\lambda^{(c)}_i=\frac{\Delta_i-1}{\Delta_i}\lambda_i$ and $\lambda^{(e)}_i=\frac{1}{\Delta_i}\lambda_i$, respectively. 

The intercell interference experienced by a typical CEU, is given  by
\begin{multline}
I=\sum\limits_{j=1}^{K}\left\{ \sum\limits_{z_c\in\theta_{j}^{(c)}}\tau(RB_{jz_c}=b)P_jg_{jz_c}r_{jz_c}^{-\alpha_j}\right\} \\+\sum\limits_{j=1}^{K}\left\{ \sum\limits_{z_e\in\theta_{j}^{(e)}}\tau(RB_{jz_e}=b)\phi_j P_jg_{jz_e}r_{jz_e}^{-\alpha_j} \right\}
\label{effectiveICI}
\end{multline}
in which $\theta_{j}^{(c)}$ and $\theta_{j}^{(e)}$ are the set of interfering BSs transmitting with a cell-center and cell-edge power level in tier $j$; $g_{jz}$ and $r_{jz}$ are the channel power gain and distance from the user to the interfering BS $z$ in the tier $j$; $\tau(RB_{jz_c}=b)$ and $\tau(RB_{jz_e}=b)$ is the indicator functions that take value 1 if the RB $b$ is 
occupied as a cell-center or cell-edge RB in cell $z_c$ or $z_e$.

When  Round Robin scheduling is deployed,  the expected values of these indicator functions are:
\begin{multline}
\mathbb{E}[\tau(RB_{jz_c}=b)]=\frac{M_{j}^{(c)}}{N_{j}^{(c)}} \text{ and } \mathbb{E}[\tau(RB_{jz_e}=b)]=\frac{M_{j}^{(e)}}{N_{j}^{(e)}} 
\end{multline} 
For optimisation purposes, the ratio of the number of RBs at a cell-edge and cell-center area is proportional to number of  the CEU and CCU \cite{optimalDelta}. Hence, we define the \textbf{optimisation factor} $\epsilon_i $ for the tier $i$ as in Equation \ref{eq:optimization} 
\begin{equation}
 \frac{M_{j}^{(c)}}{N_{j}^{(c)}}=\frac{M_{j}^{(e)}}{N_{j}^{(c)}}=\epsilon_j \quad (\forall 0<i\leq K)
\label{eq:optimization}
\end{equation} 
\\

\section{Coverage probability of a typical cell-edge user}
\label{sectionCellEdgeUserCov}
In this section, the coverage probability of a typical user that is served on a cell-edge RB, is considered. 
The CEU  is said under the coverage region if there is at least one tier $i$ among the  K tiers such that the received $SINR_{i}(\phi_i,r_i)$ from at least one BS in tier $i$ is larger than a coverage threshold $\hat{T_i}$. 
 \begin{equation}
 P^{(e)}_c=\mathbb{P}\left(\bigcup_{i=1}^K SINR_i(\phi_i,r_i) \geq \hat{T}_i \right)=\sum_{i=1}^{K}P_{Ai}P_i^{(e)}(\hat{T_i},\phi_i)
 \label{coveragedef}
 \end{equation}
where $P^{(e)}_i(\hat{T_i},\phi_i)$ is the coverage probability of the user which is associated with the tier $i$ and  $P_{Ai}$ is the probability that  the user is connected to tier $i$. When the user is under the coverage area, it can successfully decode  the received signal.  Then, it is clear that the value of $\hat{T_i}$ varies depending on the sensitivity of the user equipment. 

The received $SINR_{i}(\phi_i,r_i)$ from the tier $i$ can be expressed as a function of the distance between the user and its associated BS.
 \begin{equation}
 SINR_{i}(\phi_i,r_i)=\frac{\phi_iP_ig_{i}r_{i}^{-\alpha_i}}{I_u+\sigma^2}
 \label{SINR}
 \end{equation}
in which $g_{i}$ and $r_{i}$ is the channel power gain and distance from the user to the serving BS in tier $i$; $\phi_i$ is the transmission power ratio  that was defined in section \ref{sec:Softpara}; $\sigma^2$ is Gaussian noise.
The coverage probability in Equation \ref{coveragedef} can be rewritten as the following expression
\begin{align}
&P_c^{(e)}(\hat{T_i},\phi_i)=\int\limits_{0}^{\infty}\mathbb{P}^{(e)}_i(\hat{T_i},\phi_i|r_i)f_{R_i}(r_i)dr_i \nonumber \\
&\quad \quad =\int\limits_{0}^{\infty}\frac{2\pi\lambda_i}{P_{Ai}}r_i\exp\left(-\pi\sum\limits_{j=1}^{K}\lambda_jr_i^{\frac{2\alpha_i}{\alpha_j}} \right)\mathbb{P}^{(e)}_i(\hat{T_i},\phi_i|r_i) dr_i
\label{averageCovinter}
\end{align}
Hence, to find expression of $P_c^{(e)}(\hat{T_i},\phi_i)$, the coverage probability of the CEU at a distance $r_i$ from its serving BS, i.e. $\mathbb{P}^{(e)}_i(\hat{T_i},\phi_i|r_i)$ should be computed first.

\begin{lemma}
\label{theoremCov}
The coverage probability of the typical cell-edge user at a distance $r_i$ from its serving BS in the tier $i$ is given by
\begin{align}
&\mathbb{P}_i^{(e)}(\hat{T_i},\phi_i|r_i)=\sum\limits_{n=1}^{N_H}\frac{\omega_n}{\sqrt{\pi}}\exp\left[-\frac{\hat{T_i}r_{i}^{\alpha_i}}{\gamma(a_n)}\frac{1}{\phi_iSNR_i} \right] \nonumber \\
&\exp\left[-\pi r_i^2\sum\limits_{j=1}^{K}\epsilon_j\left(\lambda^{(c)}_jf_I(\hat{T_i},\phi_i,1,i,j) +\lambda^{(e)}_jf_I(\hat{T_i},\phi_i,\phi_j,i,j)
\right) \right]
\label{coverageConEdge}
\end{align}
where $f_I(\hat{T_i},\phi_i,\phi_j,i,j)=\\$
\scalebox{1}{$\quad \quad \quad \sum\limits_{n_1=1}^{N_H}\frac{\omega_{n_1}}{\sqrt{\pi}}\left[\frac{2\pi C(\hat{T}_i,\phi_i)^{\frac{2}{\alpha_j}}}{\alpha_j\sin\left(\frac{\pi(\alpha_j-2)}{\alpha_j} \right) } -\sum\limits_{m=1}^{N_L}\frac{c_{m}}{2}\frac{C(\hat{T}_i,\phi_i)}{c_{m}+\left(\frac{x_{m}+1}{2} \right)^{\alpha_j/2}} \right] \text{;}$
}\\
\scalebox{1}{${C(\hat{T}_i,\phi_i)=\hat{T_i}\frac{\gamma(a_{n_1})}{\gamma(a_n)}\frac{\phi_jP_j}{\phi_iP_i}r_{i}^{\alpha_i-\alpha_j}}$};
\scalebox{1}{$\gamma(a_n)=10^{(\sqrt{2}\sigma_za_n+\mu_z)/10}$};\\ \noindent{ $\omega_n$ and $a_n$,$c_m$ and $x_m$ are are weights and nodes of Gauss-Hermite, Gauss-Legendre rule respectively with order $N_{H}$}.

\textbf{\textbf{Proof}}: See Appendix
\end{lemma}
It is observed from Lemma \ref{theoremCov} that the coverage probability of a typical user is inversely proportional to exponential function of  $1/SNR$ and $r^Î±_i$ for cellular network with $\sigma^2>0$. On right-hand side of the equation, the first exponential part corresponds to the transmission SNR which is comparable with the published results and equals 1 in the case of an interference-limited network. Meanwhile, the second exponential part corresponds to the ICI which represents the relationship between one side such as density of BS, transmission ratio and optimisation factor, and other side such as coverage probability. When the FR factor $\Delta=1$, i.e. $\phi_i=\phi_j \quad \forall i,j$, then the coverage probability expression is comparable with the results in \cite{CongICC2016,flexiblecell}.
 
The main difference between the CEU and CCU is that the CEU is served by the BS transmitting at the high power level while the CCU is served by the BS transmitting with the lower power level. The coverage probability of the typical CCU can be obtained by:
\begin{align}
&\mathbb{P}_i^{(c)}(\hat{T_i},1|r_i)=\sum\limits_{n=1}^{N_H}\frac{\omega_n}{\sqrt{\pi}}\exp\left[-\frac{\hat{T_i}r_{i}^{\alpha_i}}{\gamma(a_n)}\frac{1}{SNR_i} \right] \nonumber \\
&\exp\left[-\pi r_i^2\sum\limits_{j=1}^{K}\epsilon_j\left(\lambda^{(c)}_jf_I(\hat{T_i},1,1,i,j) +\lambda^{(e)}_jf_I(\hat{T_i},1,\phi_j,i,j)
\right) \right]
\label{coverageConCenter}
\end{align} 
in which $f_{I}(\hat{T_i},1,\phi_i,i,j)$ and related symbols are defined in Lemma \ref{theoremCov}.
\begin{proposition}
\label{ProaverageCov}
The average coverage probability of the typical CEU that is randomly located in the network is given by
\begin{multline}
P^{(e)}_c=\sum\limits_{i=1}^{K}\sum\limits_{m=1}^{N_{L}}4\pi\lambda_i\frac{c_{m}(x_{m}+1)}{(1-x_{m})^3}\\
 e^{-\pi\sum\limits_{j=1}^{K}\lambda_j\left(\frac{x_{m}+1}{1-x_{m}} \right)^{\frac{2\alpha_i}{\alpha_k}}}\mathbb{P}_i^{(e)}\left(\hat{T_i},\phi_i|r_{m}=\frac{x_{m}+1}{1-x_{m}}\right)
\label{averageCov}
\end{multline}
\end{proposition}
\textbf{\textbf{Proof}} This expression of average coverage probability can be obtained from Equation \ref{averageCovinter} by using Gauss-Legendre quadrature, i.e. letting $r=\frac{t}{1-t}$.

This closed-form expression for coverage probability of a typical user in Rayleigh-Lognormal fading is simpler and more general than the previous results in \cite{Dhillon2012} which was produced for Rayleigh fading only and a FR  factor $\Delta=1$.  
\begin{proposition}
\label{limtedinter}
The average coverage probability of the cell-edge user in an interference-limited network (with no noise) the homogenous network , i.e. $\alpha_i=\alpha_j$ , $\forall 0<i,j<K$, is given by
\begin{align}
P&_c^{(e)}(\hat{T_i},\phi_i)=\sum\limits_{i=1}^{K}\sum\limits_{n=1}^{N_H} \frac{\omega_n}{\sqrt{\pi}}\nonumber \\
&\frac{1}{1+\frac{1}{\lambda_i}\sum\limits_{j=1}^{K}\epsilon_j\left(\lambda^{(c)}_jf_I(\hat{T_i},\phi_i,1,i,j) +\lambda^{(e)}_jf_I(\hat{T_i},\phi_i,\phi_j,i,j)\right)}
\label{averageCovNonoise}
\end{align}
\end{proposition}
\textbf{Proof} When $\sigma^2=0$, the desired result can be achieved  by evaluating the integral in Equation \ref{averageCovinter}.

This close-form expression is comparable to the corresponding result for Rayleigh fading and FR factor $\Delta=1$ given in \cite{HeZhuang2014}. When the number of tiers is one ($K=1$) or the densities of BSs in all tiers are the same, the average coverage probability does not depend on the density of the BS. This means the coverage probability is consistent with the changes the of number of BS \cite{Andrews}. 
\section{Average coverage probability of a typical user}
In this section, we derive the average coverage probability of a typical user (called randome user)  that is randomly located in a K-tier cellular network and can be either served as a CEU or CCU. Using the same approach that was  given in Section \ref{sectionCellEdgeUserCov}, the average coverage probability is given by
\begin{align}
P_c &= \sum\limits_{i=1}^{K}P_{Ai}P_i(\hat{T_i},\phi_i)
\end{align}
where $P_{Ai}$ is the probability that the random user is associated with the tier $i$ and  $P_i(\hat{T_i},\phi_i)$ is the average coverage probability of the random user when it is connected to the tier $i$. 

It is noticed $P_i(\hat{T_i},\phi_i)$ is  obtained by averaging the coverage probability of the user at the distance $r_i$, i.e. $\mathbb{P}_i(\hat{T_i},\phi_i|r_i)$,  over the network. Hence,
\begin{align}
P_c &=\sum\limits_{i=1}^{K}P_{Ai}\int\limits_{0}^{\infty}\mathbb{P}_i(\hat{T_i},\phi_i|r_i)f_{R_i}(r_i)dr_i \nonumber \\
&=\sum\limits_{i=1}^{K}\int\limits_{0}^{\infty}2\pi\lambda_ir_i\exp\left(-\pi\sum\limits_{j=1}^{K}\lambda_jr^{\frac{2\alpha_i}{\alpha_j}} \right) \mathbb{P}_i(\hat{T_i},\phi_i|r_i)dr_i
\label{eq:tmp1averageCovPerDis}
\end{align}
where $f_{R_i}(r_i)$ is the PDF of $r_i$ that was given in Equation \ref{PDFdistance}.

For a given tier $i$ in a LTE network enabling frequency reuse, a typical user can be served as a CEU (or CCU) when its  received SINR, i.e. $SINR_i(1,r_i)$ in Equation \ref{SINR}, is below (or above) the SINR threshold $T_i$. If the user is served as a CCU, then it is said to be under the coverage region of the tier $i$ if $SINR_i(1,r_i)$ are larger than both the $SINR$ threshold $T_i$ and the coverage threshold $\hat{T}_i$.  If the user is served as a CEU, then it can be served on different RB and transmission power. Hence, it experience new SINR, i.e. $SINR_i'(\phi_i,r_i)$,. Hence, the user in this case is under the coverage region of the tier $i$ if $SINR_i'(\phi_i,r_i)$ is above the threshold $T_i$. Subsequently, the probability $\mathbb{P}_i(\hat{T_i},\phi_i|r_i)$ of a random user is obtained by
\begin{align}
\mathbb{P}_i(\hat{T_i},\phi_i|r_i)=&\mathbb{P}(SINR'_{i}(1,r_i)>\hat{T_i})\mathbb{P}(SINR_{i}(1,r_i)>T_i)\nonumber \\
&+\mathbb{P}(SINR'(\phi_i,r_i)>\hat{T_i})\mathbb{P}(SINR_{i}(1,r_i)<T_i) \nonumber \\
=&\mathbb{P}_i^{(c)}(\hat{T_i},1|r_i)\mathbb{P}_i^{(c)}(T_i,1|r_i)\nonumber \\
&+\mathbb{P}_i^{(e)}(\hat{T_i},\phi_i|r_i)\left[1-\mathbb{P}_i^{(c)}(T_i,1|r_i) \right] 
\label{eq:tmp2averageCovPerDis}
\end{align}
in which $\mathbb{P}_i^{(e)}(\hat{T_i},\phi_i|r_i)$ and $\mathbb{P}_i^{(c)}(\hat{T_i},\phi_i|r_i)$ are given in Equation \ref{coverageConEdge} and \ref{coverageConCenter}, respectively. 

Using the same approach as in Proposition \ref{ProaverageCov}, the average coverage probability can be obtained  by

\begin{multline}
P_c=\sum\limits_{i=1}^{K}\sum\limits_{m=1}^{N_{L}}4\pi\lambda_i\frac{c_{m}(x_{m}+1)}{(1-x_{m})^3}\\
 e^{-\pi\sum\limits_{j=1}^{K}\lambda_j\left(\frac{x_{m}+1}{1-x_{m}} \right)^{\frac{2\alpha_i}{\alpha_k}}}\mathbb{P}_i\left(\hat{T_i},\phi_i|r_{m}=\frac{x_{m}+1}{1-x_{m}}\right)
\label{averageCov}
\end{multline}
In the case of interference-limited network with  a frequency reuse factor $\Delta=1$ is used, i.e. $\phi_i=\phi_j \forall 0<i,j\leq K$, then the average coverage probability is given by the same approach as in Proposition \ref{limtedinter}
\begin{dmath}
P_c=\sum\limits_{i=1}^{K}\sum\limits_{n=1}^{N_H} \frac{\omega_n}{\sqrt{\pi}}
\frac{1}{1+
	\frac{1}{\lambda_i}\sum\limits_{j=1}^{K}\lambda_j\epsilon_jf_I(\hat{T_i},1,1,i,j)}
\end{dmath}
This closed-form expression of average coverage probability is comparable with the published result in \cite{SinhCongLam2015}. The average coverage probability only depends on the optimization factor.

\section{Simulation and Discussion}
In this section, the Monte Carlo simulations are used to give a graphical verification of the accuracy of the  analytical results presented in Section 4. Furthermore, this section provides more details about the relationship between  network performance and the related parameters such as transmission power ratio, optimization factor and coverage threshold. 

Since the coverage threshold represents the signal detection probability of the user equipment, 
it does not depend on the tier in which the user is associated with, i.e. $\hat{T}_i=\hat{T}_j, \quad \forall 0<i,j<K$. To reduce the computation complexity, it is assumed that the path loss exponents at different tiers are the same. The  analytical and simulation parameters used in this paper are summarized in Table \ref{SimulationParaHeteK=2}.
\begin{table}[!h]
\centering
\begin{tabular}{|l|l|}
\hline \rule[0ex]{0pt}{3ex} Parameter & Value \\ 
\hline \rule[0ex]{0pt}{2ex} Number of tiers & $K=2$ \\ 
\hline \rule[0ex]{0pt}{2ex} Density of BSs & Tier 1, $\lambda=0.25$ \\ 
	   \rule[0ex]{0pt}{2ex} 				 & Tier 2, $\lambda=0.5$ \\ 
\hline \rule[0ex]{0pt}{2ex} Transmission power ratio between tiers  & $100$ \\ 
		\rule[0ex]{0pt}{2ex}  $(P_1/P_2)$ &  \\
\hline \rule[0ex]{0pt}{2ex} Fading channel & $\mu_z=-7.3683$ dB \\ 
		\rule[0ex]{0pt}{2ex}& $\sigma_z=8$ dB \\
\hline \rule[0ex]{0pt}{2ex} Pathloss exponent & $\alpha_1=\alpha_2=3.5$ \\ 
\hline 
\end{tabular} 
\caption{Analytical and simulation parameters}
\label{SimulationParaHeteK=2}
\end{table}

\begin{figure}[!h]
	\centering
	\includegraphics[width=1\linewidth]{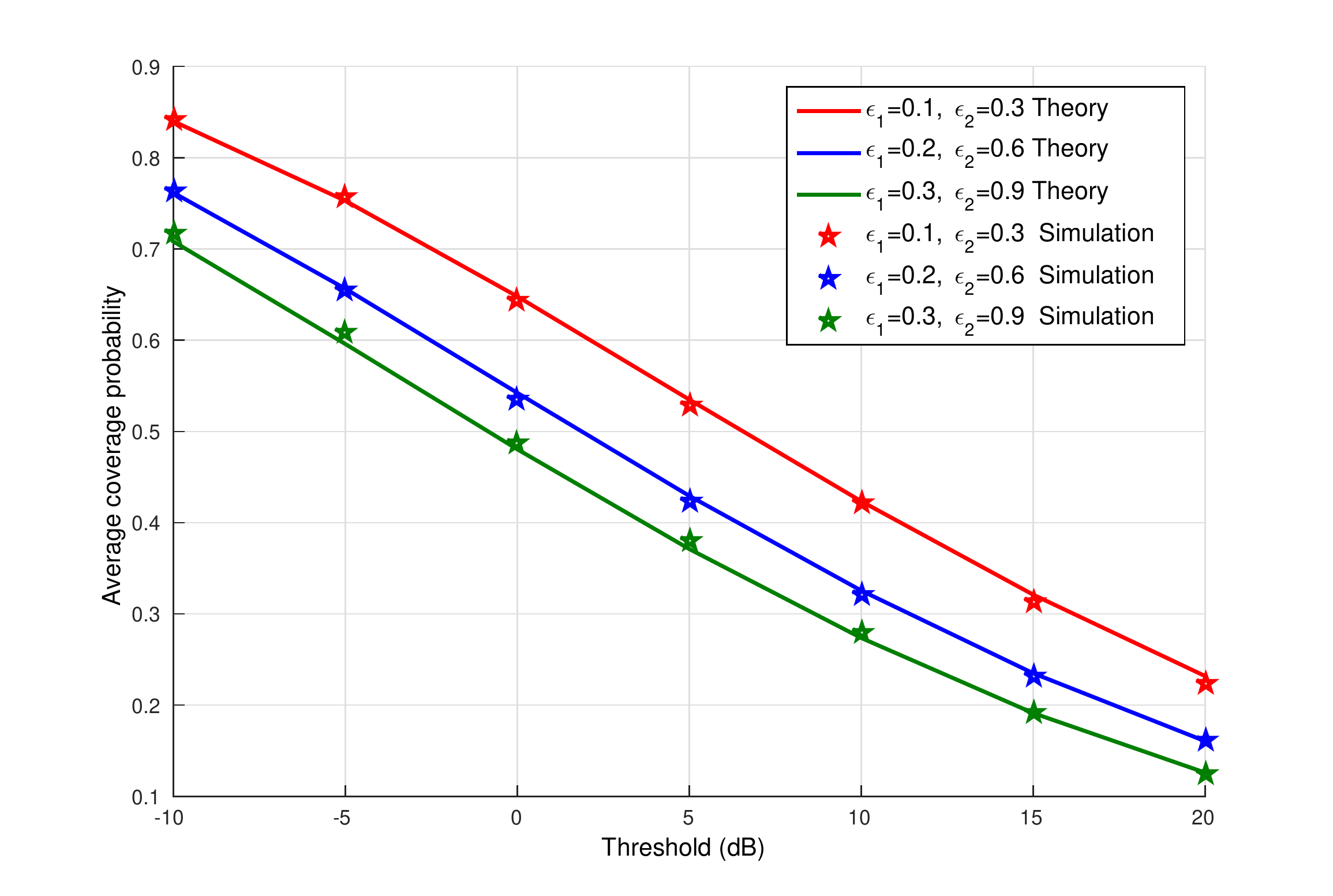}
	\caption{Coverage probability of a CEU  vs.  coverage threshold} 
	\label{fig:tmpcoverage}
\end{figure}

Figure \ref{fig:tmpcoverage} shows variation of the average coverage probability of the CEU (y-axis) versus the coverage threshold in dB (x-axis) from theory and simulation for a number of values of the optimization factors ($\epsilon_1, \epsilon_2$) which represent ratio between the number of users and RBs in each tier. 

 When the optimization factor  increases which means that there is an increase in the number of users, the average coverage probability of the CEU decreases significantly . For example, when the  coverage threshold  is $\hat{T}_{i}=0$ dB, the coverage probability decreases by 23.6\% from 0.4229 to 0.323 when the optimisation factors  doubles  from ($\epsilon_1=0.1, \epsilon_2=0.2$) to ($\epsilon_1=0.2, \epsilon_2=0.6$).

Furthermore, it is observed in Figure \ref{fig:tmpcoverage} that the average coverage probability which is  proportional to the detection probability of user equipment. When the sensitivity of user device is improved which means that the user can work with lower SINR (i.e. lower coverage threshold), the probability that  the user successfully decodes the received signals increases. Consequently, the performance of this user increases. 

\begin{figure}[!h]
\centering
\includegraphics[width=1\linewidth]{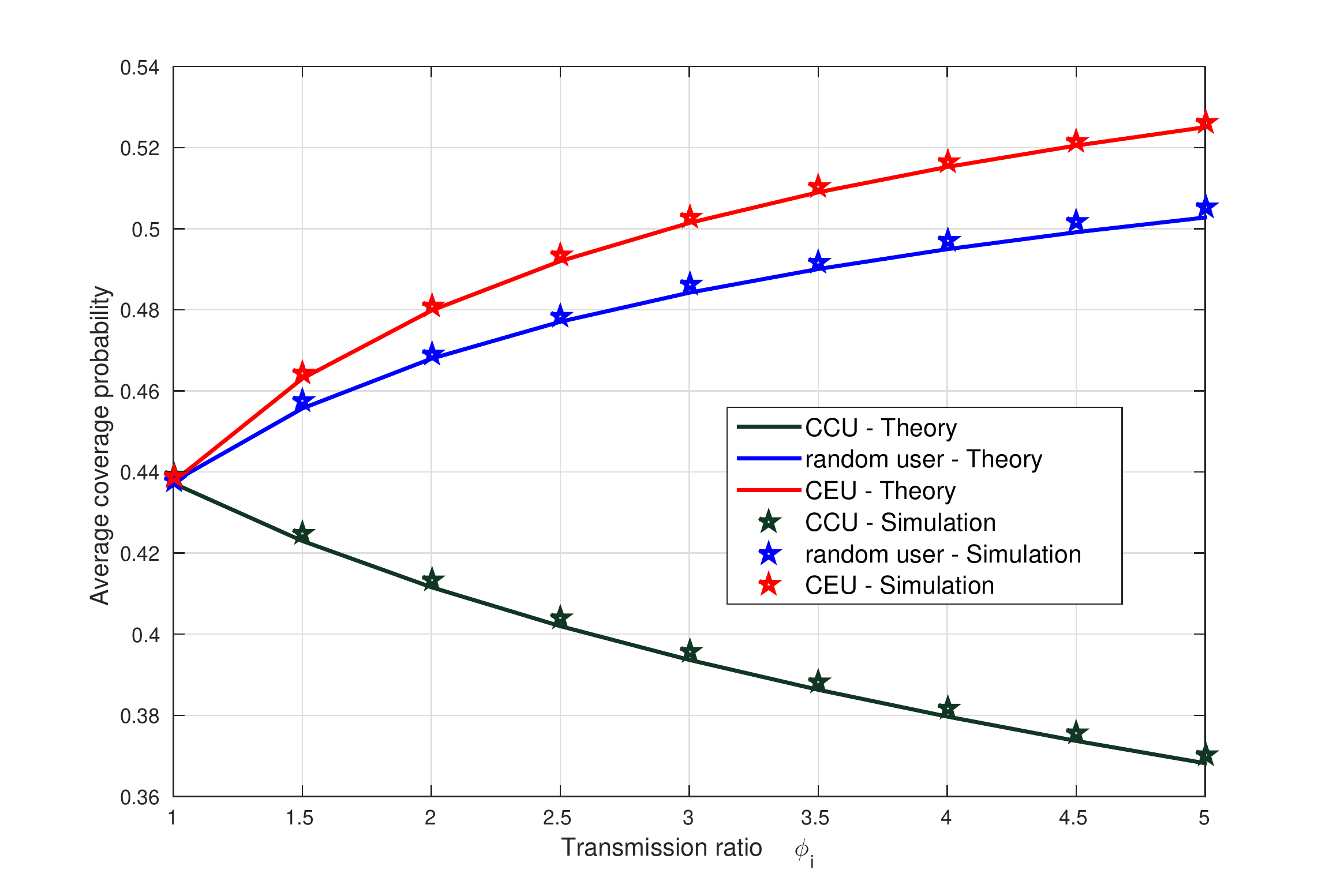}
\caption{User coverage probability versus  transmission ratio $\phi_i$}
\label{fig:3coveragepro}
\end{figure}

In Figure \ref{fig:3coveragepro}, the coverage probability of the CCU, CEU  and random user are shown  as a function of  transmission power ratio .

When the transmission power ratio increases (i.e. an increase in the transmission power on the cell-edge RB),  the received  signal power for  the CEU and the  ICI power in the network increases while the  power of the received signal of the CCU is kept constant. Hence, as shown in the figure, the coverage probability of the CEU grows moderately, while that of the CCU decreases . For the random user which can be  CCU or CEU,  the coverage probability  increases slightly. 

\section{Conclusion}
In this paper, the coverage probability of a typical user in a multi-tier network experiencing  Rayleigh-Lognormal fading is   presented through mathematical and numerical analysis. The typical user in this paper can be a CCU, CEU or random user. The main difference between the results in this paper and in other published work is that  previous works use either Strict FR or Soft FR with reuse factor $\Delta=1$ .   In this paper,  FR factor $\Delta>1$ was investigated. The results show that the coverage probability of a typical user is a function of the transmission ratio, FR factor, density of BSs and optimisation factor. Furthermore, the  results indicate that there are opposite trends between the performance of the CEU  and CCU. Therefore, the optimal problem should consider these factors together.
\section*{Appendix A}
The coverage probability in Equation \ref{coveragedef} can be rewritten as 
\begin{align}
&\mathbb{P}^{(e)}_c(\hat{T_i},\phi_i|r_i) \nonumber \\
&=\mathbb{P}(SINR_{i}(\phi_i,r_i)) \nonumber \\
 &=\sum\limits_{n=1}^{N_H}\frac{\omega_n}{\sqrt{\pi}}\mathbb{E}\left[\exp\left(- \frac{Tr_{i}^{\alpha_i}(I_u+\sigma^2) }{\phi_iP_i\gamma(a_n)}\right)  \right]  \nonumber \\
 &=\sum\limits_{n=1}^{N_H}\frac{\omega_n}{\sqrt{\pi}}\left[\exp\left(-\frac{Tr_{i}^{\alpha_i}}{\gamma(a_n)}\frac{1}{\phi_iSNR_i} \right) \mathbb{E}\left\lbrace \exp\left(- \frac{Tr_{i}^{\alpha_i}I_u}{\phi_iP_i\gamma(a_n)}\right)  \right\rbrace  \right] 
 \label{firststep}
\end{align}
in which $SNR_i=\frac{P_i}{\sigma^2}$.

By substituting Equation \ref{effectiveICI}, the expectation can be separated into two elements $\mathbb{E}^{(c)}$ and $\mathbb{E}^{(e)}$.
\begin{align*}
\mathbb{E}\left\lbrace \exp\left(- \frac{Tr_{i}^{\alpha_i}}{\phi_iP_i\gamma(a_n)}I_u\right)  \right\rbrace =  \mathbb{E}^{(c)}\text{ x }\mathbb{E}^{(e)}
\end{align*}
in which $\mathbb{E}^{(e)}$ and $\mathbb{E}^{(c)}$ are given in Equation \ref{eq:EcEe}
\begin{figure*}[!th]
\normalsize
\begin{eqnarray}
\mathbb{E}^{(c)}&=&\prod\limits_{j=1}^{K}\mathbb{E}\left\lbrace\prod\limits_{z_c\in\theta_{j}^{(c)}}\tau(RB_{jz_c}=b)\exp\left(-C(\hat{T}_i,\phi_i)g_{jz_c}\left( \frac{r_{jz_e}}{r_i}\right)^{-\alpha_j} \right)\right\rbrace \nonumber \\ 
\mathbb{E}^{(e)}&=&\prod\limits_{j=1}^{K}\mathbb{E}\left\lbrace\prod\limits_{z_e\in\theta_{j}^{(e)}}\tau(RB_{jz_e}=b)\exp\left(-C(\hat{T}_i,\phi_i) g_{jz_e}\left( \frac{r_{jz_e}}{r_i}\right)^{-\alpha_j}
\right)\right\rbrace
\label{eq:EcEe}
\end{eqnarray}
\begin{align*}
\text{in which}\quad  C(\hat{T}_i,\phi_i)=\hat{T}_i\frac{\gamma(a_{n_1})}{\gamma(a_n)}\frac{\phi_jP_j}{\phi_iP_i}r_{i}^{\alpha_i-\alpha_j}. \nonumber
\end{align*}
\hrulefill
\vspace*{4pt}
\end{figure*}

Evaluating the second group product $\mathbb{E}^{(e)}$, we have
\begin{dmath*}
\mathbb{E}^{(e)}=\prod\limits_{j=1}^{K}\mathbb{E}\left\lbrace\prod\limits_{z_e\in\theta_{j}^{(e)}}\epsilon_j\\ \quad \quad  \quad \quad  \quad \quad  \quad \mathbb{E}_{g_{jz}}\left[\exp\left(-C(\hat{T}_i,\phi_i)g_{jz_e}\left(\frac{r_{jz_e}}{r_i}  \right) ^{-\alpha_j}
\right) \right] \right\rbrace
\end{dmath*}
 Since $g_{jz_c}$ is Rayleigh-Lognormal fading channel then
\begin{align*}
=&\prod\limits_{j=1}^{K}\mathbb{E}\left\lbrace\prod\limits_{z_e\in\theta_{j}^{(c)}}\epsilon_j\sum\limits_{n_1=1}^{N_H}\frac{\omega_{n_1}}{\sqrt{\pi}}\frac{1}{1+C(\hat{T}_i,\phi_i)\left(\frac{r_{jz_e}}{r_i}  \right) ^{-\alpha_j}}
\right\rbrace\nonumber
\end{align*}
Each element of this product can be evaluated by using the properties of PPP generating function, Gamma function and Gauss-Legendre approximation quadrature shown in \cite{SinhCongLam2015}, then the expectation $\mathbb{E}^{(e)}$ equals
\begin{equation}
\mathbb{E}^{(e)}
=\exp\left( -\pi\lambda^{(e)}_jr_i^2\sum\limits_{j=1}^{K} \epsilon_jf_I(\hat{T_i},\phi_i,\phi_j,i,j)\right)
\label{eq:Ee}
\end{equation}
in which
\begin{dmath}
f_I(\hat{T_i},\phi_i,\phi_j,i,j)=
\sum\limits_{n_1=1}^{N_H}\frac{\omega_{n_1}}{\sqrt{\pi}}\left(\frac{2\pi C(\hat{T}_i,\phi_i)^{\frac{2}{\alpha_j}}}{\alpha_j\sin\left(\frac{\pi(\alpha_j-2)}{\alpha_j} \right) }\\ -\sum\limits_{m=1}^{N_{H}}\frac{c_{m}}{2}\frac{C(\hat{T}_i,\phi_i)}{c_{m}+\left(\frac{x_{m}+1}{2} \right)^{\alpha_j/2} }
\right) 
\label{eq:fi}
\end{dmath}
where  $c_m$ and $x_m$ are are weights and nodes of Gauss-Legendre rule with order $N_{L}$.
Similarly, $\mathbb{E}^{(c)}$ is achieved by
\begin{align}
\mathbb{E}^{(c)}=\exp\left( -\pi\lambda_j^{(c)}r_i^2\sum\limits_{j=1}^{K} \epsilon_j\phi_jf_I(T,\phi_i,1,i,j)\right)
\label{eq:Ec}
\end{align}
Substituting Equation \ref{eq:Ee} and \ref{eq:Ec} into Equation \ref{firststep}, the Theorem is proved.

\small{
\addcontentsline{toc}{chapter}{Bibliography}
\bibliographystyle{IEEEtran}
\bibliography{refDA}

\begin{thebibliography}{10}
\providecommand{\url}[1]{#1}
\csname url@samestyle\endcsname
\providecommand{\newblock}{\relax}
\providecommand{\bibinfo}[2]{#2}
\providecommand{\BIBentrySTDinterwordspacing}{\spaceskip=0pt\relax}
\providecommand{\BIBentryALTinterwordstretchfactor}{4}
\providecommand{\BIBentryALTinterwordspacing}{\spaceskip=\fontdimen2\font plus
\BIBentryALTinterwordstretchfactor\fontdimen3\font minus
  \fontdimen4\font\relax}
\providecommand{\BIBforeignlanguage}[2]{{%
\expandafter\ifx\csname l@#1\endcsname\relax
\typeout{** WARNING: IEEEtran.bst: No hyphenation pattern has been}%
\typeout{** loaded for the language `#1'. Using the pattern for}%
\typeout{** the default language instead.}%
\else
\language=\csname l@#1\endcsname
\fi
#2}}
\providecommand{\BIBdecl}{\relax}
\BIBdecl

\bibitem{Softdefinition}
\emph{Soft Frequency Reuse Scheme for UTRAN LTE}, 3GPP Project Document Std.
  R1-050\,507, Huawei,2005.

\bibitem{Andrews}
J.~G. Andrews, F.~Baccelli, and R.~K. Ganti, ``A new tractable model for
  cellular coverage,'' in \emph{Communication, Control, and Computing
  (Allerton), 2010 48th Annual Allerton Conference on}, Conference Proceedings,
  pp. 1204--1211.

\bibitem{ThomasDavidNovlan2011}
A.~G. J. G.~A. Thomas David~Novlan, Radha Krishna~Ganti, ``Analytical
  evaluation of fractional frequency reuse for ofdma cellular networks,''
  \emph{IEEE TRANSACTIONS ON WIRELESS COMMUNICATIONS}, vol.~10, pp. 4294--4305,
  2011.

\bibitem{Dhillon2012}
H.~S. Dhillon, R.~K. Ganti, F.~Baccelli, and J.~G. Andrews, ``Modeling and
  analysis of k-tier downlink heterogeneous cellular networks,'' \emph{Selected
  Areas in Communications, IEEE Journal on}, vol.~30, no.~3, pp. 550--560,
  2012.

\bibitem{flexiblecell}
H.-S. Jo, Y.~J. Sang, P.~Xia, and J.~Andrews, ``Heterogeneous cellular networks
  with flexible cell association: A comprehensive downlink sinr analysis,''
  \emph{Wireless Communications, IEEE Transactions on}, vol.~11, no.~10, pp.
  3484--3495, October 2012.

\bibitem{optimalFrSpectrumwithoutnoise}
W.~Bao and B.~Liang, ``Structured spectrum allocation and user association in
  heterogeneous cellular networks,'' in \emph{INFOCOM, 2014 Proceedings IEEE},
  April 2014, pp. 1069--1077.

\bibitem{SinhCongLam2015}
S.~C. Lam and K.~Sandrasegaran, ``A closed-form expression for coverage
  probability of random cellular network in composite rayleigh-lognormal fading
  channels,'' in \emph{Accepted for Publication at International
  Telecommunication Networks and Applications Conference (ITNAC)}, November,
  2015.

\bibitem{Keeler}
H.~P. Keeler, B.~Blaszczyszyn, and M.~K. Karray, ``Sinr-based k-coverage
  probability in cellular networks with arbitrary shadowing,'' in
  \emph{Information Theory Proceedings (ISIT), 2013 IEEE International
  Symposium on}, Conference Proceedings, pp. 1167--1171.

\bibitem{CongICC2016}
S.~C. Lam, K.~S. Sandrasegaran, and Q.~T. Nguyen, ``Performance of soft
  frequency reuse in random cellular networks in rayleigh-lognormal fading
  channels,'' in \emph{Accepted for publication at APCC 2016, Indonesia}.

\bibitem{DinhThiThai}
M.~Dinh Thi~Thai, C.~Lam~Sinh, T.~Nguyen~Quoc, and N.~Dinh-Thong, ``Ber of qpsk
  using mrc reception in a composite fading environment,'' in
  \emph{Communications and Information Technologies (ISCIT), 2012 International
  Symposium on}, Conference Proceedings, pp. 486--491.

\bibitem{optimalDelta}
L.~Chen and D.~Yuan, ``Generalized frequency reuse schemes for ofdma networks:
  Optimization and comparison,'' in \emph{Vehicular Technology Conference (VTC
  2010-Spring), 2010 IEEE 71st}, May 2010, pp. 1--5.

\bibitem{HeZhuang2014}
T.~O. He~Zhuang, ``A model based on poisson point process for downlink k tiers
  fractional frequency reuse heterogeneous networks,'' \emph{Physical
  Communication}, vol. Volume 13, Part B, no. Special Issue on Heterogeneous
  and Small Cell Networks, p. 3–12, 2014.

\end{thebibliography}
}

\end{document}